\documentclass[aps,prd,amsmath,amssymb,preprintnumbers,nofootinbib,a4paper,11pt]{revtex4}
\pdfoutput=1
\usepackage{amsthm}
\usepackage{graphicx}
	\graphicspath{{./NewFig/}}
\usepackage{url}
\usepackage[bookmarks, pagebackref=false]{hyperref}
\usepackage{color}
	\definecolor{rossoCP3}{cmyk}{0,.88,.77,.40}
		\definecolor{graa}{rgb}{0.8,0.8,0.8}
		\definecolor{blaa}{rgb}{0.2,0.2,0.6}
		\hypersetup{
			colorlinks, 
			bookmarksopen, 
			bookmarksnumbered,
			citecolor=blaa, 		
			linkcolor=rossoCP3,	
			urlcolor=rossoCP3,			
			}
\usepackage{dcolumn}
\usepackage{bm}
\usepackage{bbm}
\usepackage{pxfonts}

\usepackage{epsfig}
\usepackage{placeins}

\usepackage[ margin=5pt, font=small,labelfont=bf, justification=raggedright]{caption}
\usepackage{youngtab}
\usepackage{slashed}
\usepackage{subfig}

\newcommand{\beq}{\begin{eqnarray}}
\newcommand{\eeq}{\end{eqnarray}}

\newcommand{\bmp}{\noindent\begin{minipage}{16cm}}
\newcommand{\emp}{\end{minipage}\vskip 7mm} 

\def\lsim{\mathrel{\rlap{\lower4pt\hbox{\hskip1pt$\sim$}}
    \raise1pt\hbox{$<$}}}                
\def\gsim{\mathrel{\rlap{\lower4pt\hbox{\hskip1pt$\sim$}}
    \raise1pt\hbox{$>$}}}                

\newcommand{\drawsquare}[2]{\hbox{%
\rule{#2pt}{#1pt}\hskip-#2pt
\rule{#1pt}{#2pt}\hskip-#1pt
\rule[#1pt]{#1pt}{#2pt}}\rule[#1pt]{#2pt}{#2pt}\hskip-#2pt
\rule{#2pt}{#1pt}}

\newcommand{\Yfund}{\raisebox{-.5pt}{\drawsquare{6.5}{0.4}}}

\begin{document}

\title{\LARGE \color{rossoCP3} Higgs Discovery: Impact on Composite Dynamics \\{ \rm  \large Technicolor \& eXtreme Compositeness} \\{\it Thinking Fast and Slow\footnote{Proceedings for the plenary talks delivered at the {\it X$_{th}$ Quark Confinement and the Hadron Spectrum} conference in Munich, Germany and the {\it Strong Coupling Gauge Theories in the LHC Perspective} meeting in Nagoya, Japan}}}
 \author{Francesco Sannino}\email{sannino@cp3.dias.sdu.dk} 
  \affiliation{
{ \color{rossoCP3}  \rm CP}$^{\color{rossoCP3} \bf 3}${\color{rossoCP3}\rm-Origins} \& the Danish Institute for Advanced Study {\color{rossoCP3} \rm DIAS},\\ 
University of Southern Denmark, Campusvej 55, DK-5230 Odense M, Denmark.
}

\begin{abstract}
I discuss the impact of the discovery of a Higgs-like state on composite dynamics starting by critically examining the reasons in favour of either an elementary or composite nature of this state. Accepting the standard model interpretation I re-address the standard model vacuum stability within a Weyl-consistent computation. 

I will carefully examine the fundamental reasons why what has been discovered might not be the standard model Higgs. Dynamical electroweak breaking naturally addresses a number of the fundamental  issues unsolved by the standard model interpretation. However this paradigm has been challenged by the discovery of a not-so-heavy Higgs-like state. I will therefore review the recent discovery \cite{Foadi:2012bb} that the standard model top-induced
radiative corrections naturally reduce the intrinsic non-perturbative mass of the composite Higgs state towards the desired experimental value. Not only we have a natural and testable working framework but we have also suggested specific gauge theories  that can realise, at the fundamental level,  these minimal models of dynamical electroweak symmetry breaking.   These strongly coupled gauge theories are now being heavily investigated via first principle lattice simulations with encouraging results. 

The new findings show that the recent naive claims made about new strong dynamics at the electroweak scale being disfavoured by the discovery of a not-so-heavy composite Higgs are unwarranted.

I will then introduce the more speculative idea of extreme compositeness according to which not only the Higgs sector of the standard model is composite but also quarks and leptons, and provide a toy example in the form of gauge-gauge duality.
 \vskip .1cm
{\footnotesize  \it Preprint: CP$^3$-Origins-2012-024  DNRF 90\& DIAS-2012-24}
 \end{abstract}

\maketitle

\newpage
     
\section{Thinking fast and slow}

\begin{figure}[b]
\begin{center} \includegraphics[width=.9\textwidth]{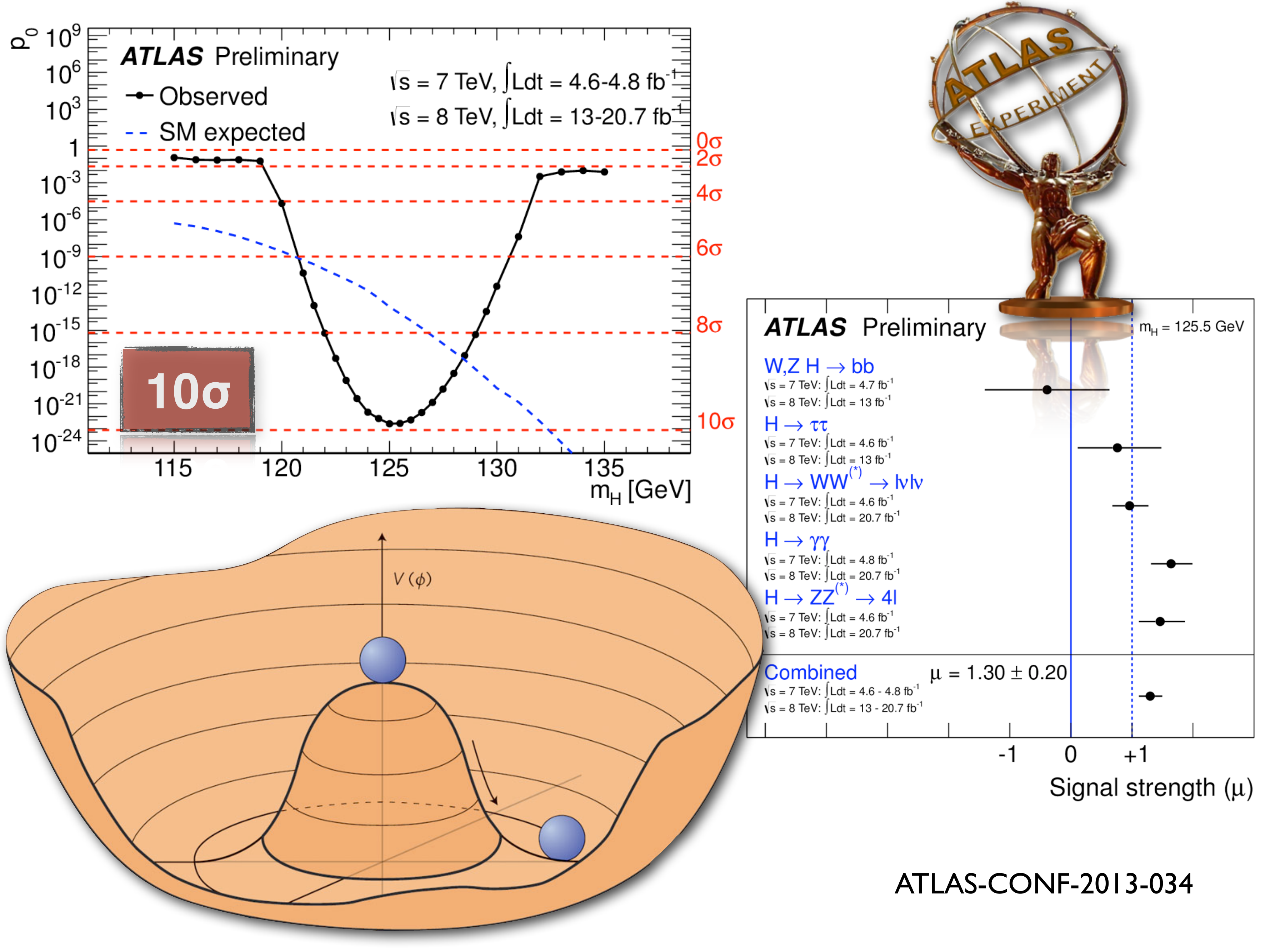} \end{center}
\caption{Top left plot is the local probability p0 for a background-only experiment to be more signal-like than the observation as a function of the mass of the Higgs or the combination of all channels. The dashed curve shows the median expected local p0 under the hypothesis of a standard model Higgs boson production at that mass.
The horizontal dashed lines indicate the p values corresponding to signiÞcances from 0 to 10. Right bottom plot  reports the measurements of the signal strength parameter for  a mass of the Higgs of 125.5 GeV for the individual channels and their combination. The results are from \cite{ATLAS:2013sla}. The mexican hat potential represents the standard model Higgs potential.}\label{SMe}
\end{figure}

Excitingly, both the ATLAS and CMS collaborations at the Large Hadron Collider (LHC) have independently reported the discovery of a new particle \cite{:2012gk,:2012gu} with properties consistent with the standard model (SM) Higgs. The current status is summarised in Fig.~\ref{SMe}. 
  The  burning question is  whether the new particle state is indeed the SM Higgs, given that the current experimental status does not allow for a precise determination of all its properties. 

\subsection{Thinking fast}

\begin{figure}[b]
\begin{center} \includegraphics[width=.8\textwidth]{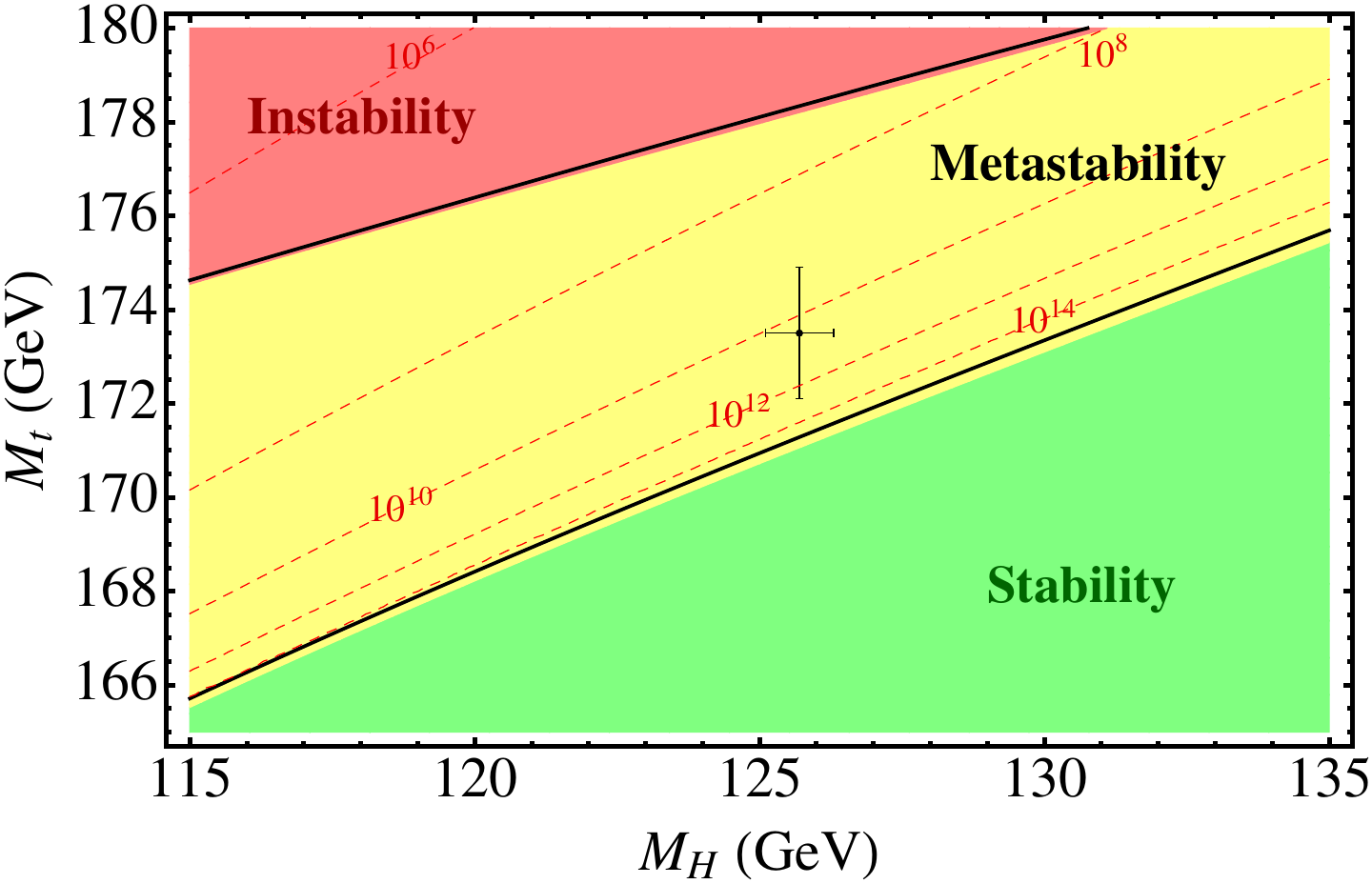} \end{center}
\caption{Standard model stability analysis based on the effective standard model Higgs quartic coupling and respecting the Weyl consistency conditions \cite{Antipin:2013sga}. The red region indicates instability, the yellow metastability and the green absolute stability following the analysis presented in \cite{Antipin:2013sga}. For comparison, the black lines indicate the bounds from \cite{Degrassi:2012ry}. The point with error bars shows the experimental values of the top \cite{Beringer:1900zz} and Higgs \cite{CMS-PAS-HIG-13-005} masses. The red dashed lines show the value in GeV at which the Higgs effective self-coupling crosses zero.}\label{stability}
\end{figure}

It is tempting to accept the simplest paradigm, i.e. the SM Higgs, for the following reasons: 
\begin{itemize}
\item It is the most minimal renormalizable model one can write able to break the electroweak symmetry preserving the $SU(2)_c$ custodial symmetry and compatible with all the experimental data.
\item It gives masses to the SM fermions, including the SM neutrinos \footnote{Although the SM Higgs mechanism does not require the presence of right-handed neutrinos, at the same time, it does not forbid their existence. Henceforth a Dirac mass term is easy to add in the SM. A direct Majorana mass term for the SM left-handed neutrinos requires the existence of new physics since there are no renormalizable operators within the SM able to accommodate it. However, to date, there is no experimental evidence favouring a Majorana mass term. Furthermore, once one has accepted the fine-tuning in the Higgs sector the quark and lepton Yukawa couplings can be fine-tuned too.}, while enacting the largest number of symmetries protecting against unobserved flavour changing neutral currents.
 \end{itemize}
If one accepts the SM Higgs paradigm to be valid it becomes imperative to investigate its vacuum stability \cite{Degrassi:2012ry,Antipin:2013sga} against quantum corrections. Because at high energies the SM displays classical conformality, the SM quantum corrections must satisfy the Weyl consistency conditions taken into account for the first time in \cite{Antipin:2013sga,Antipin:2013pya}. These conditions help organise the SM perturbative series and the resulting stability plot is shown in Fig.~\ref{stability}. 
The line of absolute stability, in the top-mass versus Higgs mass plane, is within two sigma from the SM point with the largest uncertainty coming from the measurement of the top mass  \cite{Antipin:2013sga}. The next-to-leading Weyl consistent analysis is therefore relevant and should be performed as suggested in \cite{Antipin:2013sga}. However according to this analysis the SM might very well be in a metastable state to tunnel to the true groundstate  located at much higher values of the Higgs field. The unity probability that this occurs within the age of the Universe defines the boundary between the stable and metastable region. 

\subsection{Thinking slow}
Accepting the SM Higgs paradigm does not come cheap and following Daniel Kahneman\footnote{Winner of the Nobel Prize in Economics and the author of the book {\it Thinking Fast and Slow}.} this is the time to think slow and examine the reasons why what has been discovered might not be the SM Higgs.

\begin{itemize}
\item The SM Higgs {\it does not explain} spontaneous symmetry breaking, at best parametrizes the phenomenon. 

\item Scalars are not fundamental representations of the Lorentz group, spin one-half fermions are. No fundamental (pseudo)scalar has ever been discovered so far in Nature. It  would be the most important discovery made at the LHC \footnote{ By linking scalars degrees of freedom to fermions, which are fundamental representations of the Lorentz group, supersymmetry \cite{Dimopoulos:1981zb,Dimopoulos:1981yj}  apparently gives a natural way to understand why scalars (and by supersymmetry also pseudoscalars) might also be natural, However, supersymmetry itself might be an emergent phenomenon \cite{Antipin:2011ny}. {\it Why would Nature possess such a large symmetry and then hide it so well?}}.

\item The masses of the SM fermions are not explained but fitted to their value. The same is true for any other extension of the SM with fundamental scalars. 

\item The SM Higgs sector alone cannot explain why we observe at least three matter generations. 

\item The dark matter problem is unexplained in the SM, although it is straightforward to construct extensions of the SM featuring perturbative or non-perturbative dark matter components. 

\item There is no explanation for the strong CP problem. 

\item There is no working mechanism for baryogenesis and dark matter genesis, if asymmetric. 

\end{itemize} 
I have not considered the grand unification problem of the SM gauge couplings since, although interesting, requires a number of ad hoc theoretical assumptions such as, for example, single step unification.  There is no reason to believe that all the couplings must unify simultaneously. In fact, we already assume, given that we do not know how to quantise gravity, that gravity unifies at a higher energy scale. 

A composite Higgs and associate composite sector represent a natural solution to several of the above mentioned issues \footnote{By composite, I mean composite by four-dimensional fermionic matter in the form of a strongly coupled gauge theory. One can, of course, enlarge the space of theories or the idea of compositeness, but should, at the same time, declare the SM problems is set to solve. For example, simply writing an effective Lagrangian to replace the Higgs sector does not  imply that the new Higgs sector is made by more fundamental matter. In fact, underlying purely bosonic nonsupersymmetric theories admit low energy effective theories describing the associated Goldstone excitations. If this happens  the first three problems of the list are not addressed.  }. 
  The time-honoured example of composite dynamics is Technicolor (TC) \cite{Weinberg:1979bn,Susskind:1978ms}.  In TC the Higgs sector of the Standard Model is replaced by a new gauge dynamics featuring fermionic matter.  
  
  Supersymmetry and TC require both the introduction of yet another sector. For supersymmetry it is the supersymmetry breaking sector and for TC one needs the new sector to account for the masses of the SM fermions.  

Additional exotic possibilities have been envisioned by theorists and are actively pursued in experiments.
For a complete overview and status of different extensions of the SM, we refer to the following recent reviews \cite{Buchmueller:2011ab,Altarelli:2012dq,Barbieri:2012sk}.

\section{Light scalar from a heavy composite state: The power of radiative corrections}

If new strong dynamics ~\cite{Weinberg:1975gm,Dimopoulos:1979es,Eichten:1979ah,Sannino:2008ha,Sannino:2009za} underlies the Higgs mechanism, is it possible for the corresponding spectrum to feature a 125 GeV composite scalar? To answer this question one must disentangle the SM radiative corrections from the dynamical mass $M_H^{TC}$, {\em i.e.} the mass stemming purely from composite dynamics. Because of the large and negative radiative corrections from the top loop, we argued in \cite{Foadi:2012bb} that, for a SM-like top-Yukawa coupling, the dynamical mass of the scalar required to match the observations is of the order of $M_H^{TC}\sim 600/\sqrt{N_{\rm TD}}$ GeV, where $N_{\rm TD}$ is the number of weak technidoublets. This is a significant increase in the value of the dynamical mass compared to the observed 125 GeV value, often incorrectly identified with the dynamical mass.  Additionally, if the dominant decay channels are into SM states, then for a physical mass of 125 GeV the TC Higgs is narrow simply because of kinematics. A famous example in strong dynamics is the $f_0(980)$ resonance, which is extremely narrow because the decay mode into $K\bar{K}$ is below threshold \cite{Harada:1995dc,Weinberg:2013cfa}.

To make the point above explicit consider TC theories (or any other new similar strong dynamics) featuring, at scales below the mass of the technirho $M_{\rho}$, only the eaten Goldstone bosons and the TC Higgs $H$.  We assume the TC dynamics to respect $SU(2)_c$ custodial isospin symmetry, and adopt a nonlinear realization for the composite states. The latter are thus classified according to linear multiplets of $SU(2)_c$: the electroweak Goldstone bosons $\pi^a$, with $a=1,2,3$, form an $SU(2)_c$ triplet, whereas the TC Higgs is an $SU(2)_c$ singlet. The elementary SM fields are multiplets of the electroweak group. The Yukawa interactions of the TC Higgs with SM fermions are induced by interactions beyond the TC theory itself, {\em e.g.} extended technicolor (ETC)~\cite{Dimopoulos:1979es,Eichten:1979ah}.

\begin{figure}
\begin{center} \includegraphics[width=10cm]{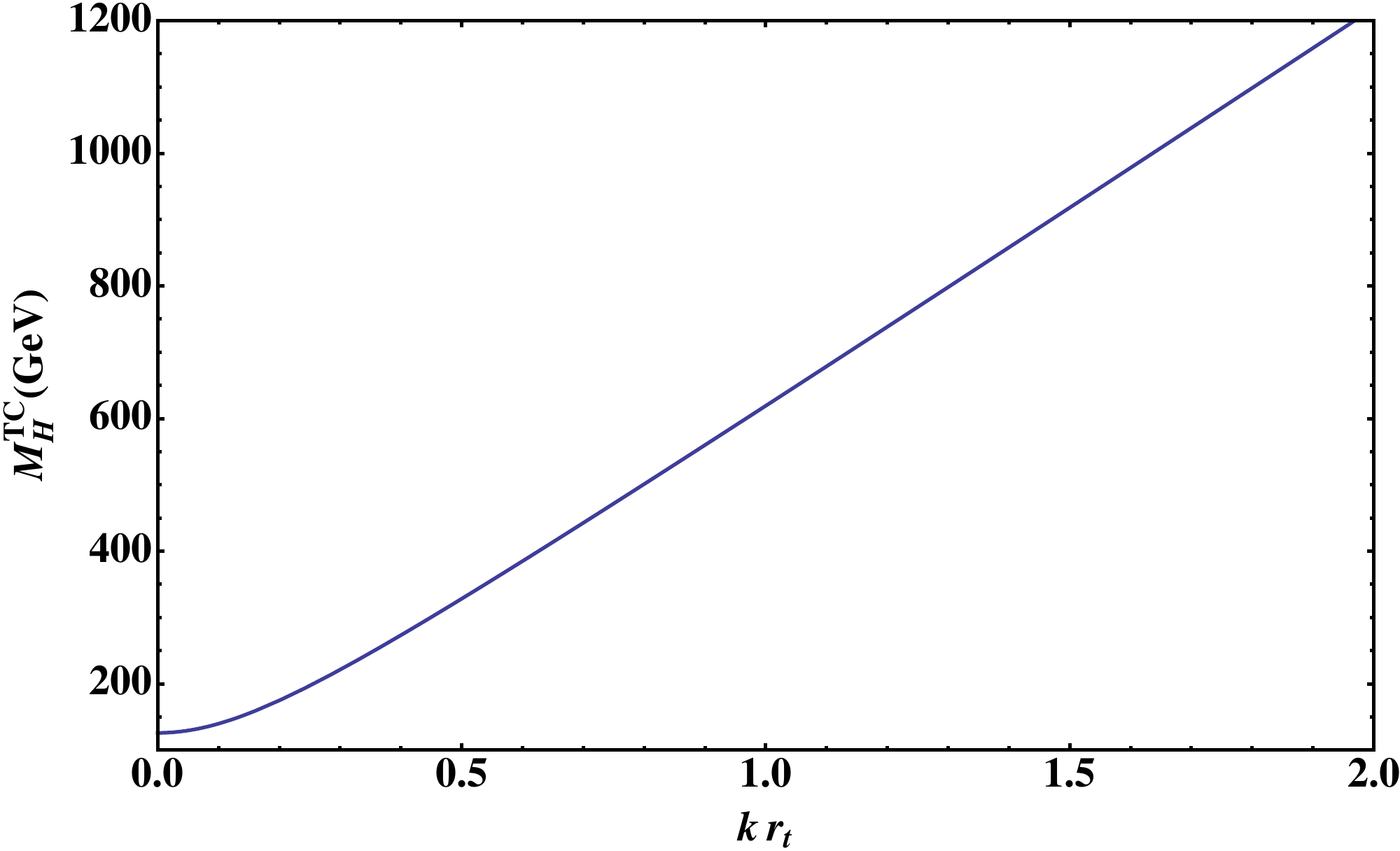}\end{center}
\caption{Mass of the TC Higgs as a function of the product $\kappa \, r_t$, using the formula $M_H^{TC} = \sqrt{M^2_H + 12 \kappa^2 r_t^2 m_t^2}$.}
\label{krt2}
\end{figure}

Assuming that the only non-negligible sources of custodial isospin violation are due to the Yukawa interactions, and retaining only the leading order terms in a momentum expansions, leads to the effective Lagrangian
\begin{eqnarray}
{\cal L} &=& {\cal L}_{\overline{\rm SM}}
+\left(1+\frac{2 r_\pi}{v}H+\frac{s_\pi}{v^2}H^2\right)\frac{v^2}{4}{\rm Tr}\ D_\mu U^\dagger D^\mu U
+ \frac{1}{2}\ \partial_\mu H\ \partial^\mu H - V[H]
 \nonumber \\
&-&
\ m_t\left(1+\frac{r_t}{v}H\right)
\Bigg[\overline{q}_L\ U\ \Bigg(\frac{1}{2}+T^3\Bigg)\ q_R + {\rm h.c.} \Bigg] 
- m_b\left(1+\frac{r_b}{v}H\right)
\Bigg[\overline{q}_L\ U\ \Bigg(\frac{1}{2}-T^3\Bigg) \ q_R + {\rm h.c.} \Bigg] + \cdots \nonumber \\
&-&\Delta S\ W^a_{\mu\nu}B^{\mu\nu}\ {\rm Tr}\  T^a U T^3 U^\dagger +{\cal O}\left(\frac{1}{M_\rho}\right)
\label{eq:L}
\end{eqnarray}
where $ {\cal L}_{\overline{\rm SM}}$ is the SM Lagrangian without Higgs and Yukawa sectors, the ellipses denote Yukawa interactions for SM fermions other than the top-bottom doublet $q\equiv(t,b)$, and ${\cal O}(1/M_\rho)$ includes higher-dimensional operators, which are suppressed by powers of $1/M_\rho$. In this Lagrangian $v\simeq 246$ GeV is the electroweak vev, $U$ is the usual exponential map of the Goldstone bosons produced by the breaking of the electroweak symmetry, $U=\exp\Big(i 2 \pi^a T^a/v\Big)$, with covariant derivative $D_\mu U\equiv \partial_\mu U -i g W^a_\mu T^a U + i g^\prime U B_\mu T^3$, $2T^a$ are the Pauli matrices, with $a=1,2,3$, and $V[H]$ is the TC Higgs potential. $\Delta S$ is the contribution to the $S$ parameter from the physics at the cutoff scale, and is assumed to vanish in the $M_\rho\to\infty$ limit. The interactions contributing to the Higgs self-energy are
\begin{eqnarray}
{\cal L}_H &\supset&  \frac{2\ m_W^2\ r_\pi}{v}\ H\ W^+_\mu\ W^{-\mu}
+\frac{m_Z^2\ r_\pi}{v}\ H\ Z_\mu\ Z^\mu - \frac{m_t\ r_t}{v}\ H\ \bar{t}\ t \nonumber \\
&+& \frac{m_W^2\ s_\pi}{v^2}\ H^2\ W^+_\mu\ W^{-\mu}
+ \frac{m_Z^2\ s_\pi}{2\ v^2}\ H^2\ Z_\mu\ Z^\mu\ .
\label{eq:hLagr}
\end{eqnarray}
The tree-level SM is recovered for
$r_\pi =s_\pi=r_t=r_b=1$.
%
%
%
%
%
%
%
%
\begin{figure}
\begin{center} \includegraphics[width=5.0in]{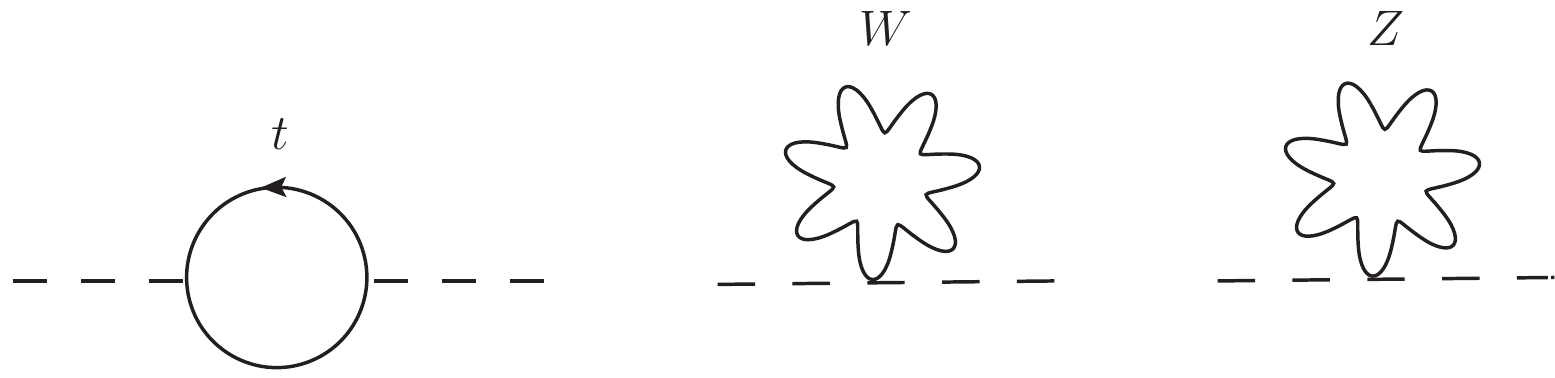} \end{center}
\caption{Quadratically divergent diagrams contributing to the Higgs mass.
The gauge boson exchanges are computed in Landau gauge: then the seagull diagrams, with a single $W$ and $Z$ exchange, are the only quadratically divergent one-loop diagrams with gauge boson exchanges.}
\label{fig:diagrams}
\end{figure}
We divide the radiative corrections to the TC Higgs mass into two classes: external contributions, corresponding to loop corrections involving elementary SM fields, and TC contributions, corresponding to loop corrections involving TC composites only. The latter contribute to the dynamical mass $M_H^{TC}$ which has been estimated in \cite{Foadi:2012bb}. In order to isolate the SM contributions we work in Landau gauge. Here transversely polarized gauge boson propagators correspond to elementary fields, and massless Goldstone boson propagators correspond to TC composites \cite{Foadi:2012ga}. The only SM contributions to the TC Higgs mass which are quadratically divergent  in the cutoff   come from the diagrams of Fig.~\ref{fig:diagrams}. Retaining only the quadratically divergent terms leads to a physical mass $M_H$ given by
\begin{eqnarray}
M_H^2=(M_H^{TC})^2 + \frac{3 (4\pi \kappa F_{\Pi})^2 }{16\pi^2 v^2}
\left[-4 r_t^2 m_t^2 +2 s_\pi \left(m_W^2+\frac{m_Z^2}{2} \right) \right] + \Delta_{M_H^2}(4\pi \kappa F_{\Pi})\ ,
\label{quadcorMrho}
\end{eqnarray}
where $F_\Pi$ is the TC-pion decay constant, and $\Delta_{M_H^2}(4\pi \kappa F_{\Pi})$ is the counterterm. The cutoff is estimated to be $4\pi\kappa F_\pi$, where $\kappa$ is a number of order one. The latter scales like $1/\sqrt{d(R_{\rm TC})}$ if the cutoff is identified with the technirho mass, or is a constant if the cutoff is of the order of $4\pi F_{\Pi}$. Provided $r_t$ is also of order one, the dominant radiative correction is due to the top quark. For instance, if $F_\Pi=v$, which is appropriate for a TC theory with one weak technidoublet, then $\delta M_H^2 \sim - 12 \kappa^2 r_t^2 m_t^2 \sim -\kappa^2 r_t^2 (600 \, {\rm GeV})^2$. In Fig.~\ref{krt2} we plot the mass of the TC Higgs as a function of the product $\kappa \, r_t$ using the formula $M_H^{TC} = \sqrt{M^2_H + 12 \kappa^2 r_t^2 m_t^2}$. This is obtained in the simple approximation of neglecting the weak gauge boson contributions, and having set to zero the counterterm in~(\ref{quadcorMrho}). This shows that the dynamical mass of the TC Higgs can be substantially heavier than the physical mass, $M_H\simeq 125$ GeV.

We have also argued that there exist a large number of underling fundamental theories available \cite{Sannino:2004qp,Hong:2004td,Christensen:2005cb,Dietrich:2006cm,Sannino:2009aw} to construct minimal models of dynamical electroweak symmetry breaking. Minimal refers to the fact that we have only a doublet of QCD neutral techni-fermions gauged under the electroweak symmetry featuring a gauge dynamics drastically different from a scaled up version of QCD.

Furthermore some of these theories can feature a scalar singlet with dynamical mass $M_H^{TC}\sim 600/\sqrt{N_{\rm TD}}$ GeV \cite{Sannino:2008ha,Foadi:2012bb}\footnote{Furthermore, reduction of the TC-Higgs dynamical mass may originate from walking (or near-conformal) dynamics~\cite{Holdom:1981rm}, both for the fundamental or for higher-dimensional representations. Walking dynamics is useful to alleviate the tension with flavor changing neutral currents, and to reduce the value of the $S$-parameter~\cite{Appelquist:1998xf,Kurachi:2006ej}. The latter, however, is not expected to vanish in conformal field theories~\cite{Sannino:2010ca,Sannino:2010fh,DiChiara:2010xb}, as well as in TC theories featuring near-conformal dynamics~\cite{Foadi:2012ga}. In the literature a light TC Higgs originating from walking-type dynamics is also known as {\em technidilaton}~\cite{Yamawaki:1985zg,Bando:1986bg,Dietrich:2005jn,Appelquist:2010gy}.}. These theories are being investigated via first principle lattice simulations with promising results \cite{Catterall:2007yx,Catterall:2008qk,Hietanen:2008mr,Catterall:2009sb,DelDebbio:2010hx,Kogut:2010cz,Kogut:2011bd,Bursa:2011ru,Giedt:2012rj} with interesting results for the physical spectrum \cite{Fodor:2012ty,Lewis:2011zb,Hietanen:2012sz}. More recently we have also been investigating on the lattice new classes of models featuring interesting patterns of chiral symmetry breaking \cite{Lewis:2011zb,Hietanen:2012sz} leading to highly interesting models of both light and heavy composite dark matter \cite{Gudnason:2006yj,Ryttov:2008xe,Foadi:2008qv,Frandsen:2009mi,DelNobile:2011je} as well as asymmetric dark matter genesis \cite{Gudnason:2006yj,Ryttov:2008xe,Frandsen:2009mi,Belyaev:2010kp}. The phenomenology associated to minimal models of dynamical electroweak symmetry breaking is summarised in \cite{Appelquist:1998xf,Foadi:2007ue,Belyaev:2008yj,Andersen:2011yj,Franzosi:2012ih}.  

Furthermore  the couplings to $WW$ and $ZZ$ of the composite Higgs are naturally of the order of the SM Higgs in minimal models of composite dynamics as explained in \cite{Foadi:2012bb}. 

The results \cite{Foadi:2012bb} reviewed above clearly demonstrate that several recent and highly naive claims made about new strong dynamics at the electroweak scale being disfavoured by the discovery of a not so heavy composite Higgs are unwarranted.

\section{eXtreme compositeness: An alternative paradigm \& a toy model}

Often in TC-inspired models, or any other composite Higgs theory, only the Higgs sector, and perhaps part of the flavour sector, are imagined to be made of something else. This theory is expected to be strongly coupled at the electroweak theory but cannot be directly used to give masses to the SM fermions. In fact in these frameworks the flavour sector requires extra strong dynamics  \cite{Ryttov:2011my,Ryttov:2010fu,Ryttov:2010jt,Ryttov:2010kc,Appelquist:2003hn}.  On the other hand it is quite natural to expect that the flavour and the Higgs sector are intimately related \cite{Fukano:2010yv} and that a better understanding would require an unified description of these sectors. 

Extreme compositeness aims, indeed, to a deeper understanding of the interplay between these two sectors. There might be many ways in which this relation can manifest itself, here we explore the possibility that extreme compositeness manifests itself in the form of electro-magnetic duality at the SM model level \cite{Sannino:2011mr}. 

According to this idea of extreme compositeness the SM is currently written in terms of {\it magnetic} fields and the dual {\it electric} description contains, besides gauge bosons, features only fermionic degrees of freedom \cite{Sannino:2011mr}. We will elucidate our idea using  the investigations on gauge-gauge duality presented in \cite{Sannino:2009qc,Sannino:2009me,Sannino:2010fh,Mojaza:2011rw} and already used for relevant phenomenological predictions in \cite{Sannino:2010fh}.  The gauge-gauge duality framework was pioneered in a series of  ground breaking papers by Seiberg  \cite{Seiberg:1994bz,Seiberg:1994pq} for supersymmetric gauge theories. 

Quarks, leptons and the Higgs itself are degrees of freedom which are to be considered elementary in the magnetic description, but arise as composite states in the dual electric variables. The reason why it has been natural to introduce these states first when introducing the SM  is that the magnetic description is perturbative at the electroweak energy scale while the electric one is strongly coupled rendering harder the identification with the electrical variables. We will therefore make use of the logical possibility  that asymptotically free gauge theories have magnetic duals. 

Experimentally three generations of quarks and leptons have been discovered with the left handed fields transforming according to the doublet representation of the weak interactions and the right handed fields are singlets with respect to the weak interactions. Electromagnetic interactions are felt by left and right transforming electrically charged states. The Higgs sector of the SM is being experimentally  tested. To illustrate how the basic idea works I start with a toy model featuring a more symmetric looking toy SM matter content summarized in Table~\ref{SM}. 
\begin{table}[h]
\[ \begin{array}{|c| c | c c | } \hline
{\rm Fields} &  \left[ SU(3) \right] & SU(N_f)_L &SU(N_f)_R   \\ \hline \hline
q &\Yfund &{\overline{\Yfund }}&1 \\
\widetilde{q} & \overline{\Yfund}&1 &  {\Yfund}\\
l &1 &{\overline{\Yfund}}&1 \\
\widetilde{l} & 1&1 &  {\Yfund} \\
{H} & 1&\Yfund &  \overline{\Yfund}\\
 \hline \end{array} 
\]
\caption{I summarize here the SM fermionic matter content. Here $H$ is a generalized Higgs field. $SU(3)$ is the color gauge group and $N_f = 2n_g$ with $n_g$ the number of fermion generations.}
\label{SM}
\end{table}

With $q_{\alpha,{c}}^i$ I indicate the two component left spinor where $\alpha=1,2$
is the spin index, $c=1,...,3$ is the color index while
$i=1,...,N_f$ represents the flavor. $\widetilde{q}^{\alpha,c}_i$
is the two component conjugated right spinor. Similarly the leptonic fields are summarized in the table with $l_{\alpha}^i$ the two component left spinor and $\widetilde{l}^{\alpha,c}_i$ the two component conjugated right spinor. Here $N_f = 2 n_g$ with $n_g$ the number of SM generations. The weak interactions are naturally embedded within the flavor group $SU(2n_g)_L \times SU(2n_g)_R$ by opportunely gauging $n_g$ times the $SU(2)_L \times U(1)_Y$ subgroup. The generalized Higgs field has been chosen to transform according to the bifundamental representation of $SU(2n_g)_L \times SU(2n_g)_R$ and therefore is not the minimal choice but it is the most natural one here. We set $SU(3)$ in between square brackets  in the table to indicate that this is the gauge group we are concentrating on to discuss how our gauge-gauge duality may work.  We therefore switch off the weak interactions for the time being.  {}From Table~\ref{SM} one is naturally led to consider the leptons as the forth color of an extended color group $ \left[ SU(4) \right] $. This is the renowned Pati-Salam  \cite{Pati:1973rp,Pati:1973uk,Pati:1974yy} extension of the SM generalized to $N_f /2 = n_g$ generations. In Table~\ref{PS}  the spectrum of the SM is summarized with respect to the Pati-Salam $SU(4)$  gauge group. 
\begin{table}[h]
\[ \begin{array}{|c| c | c c   | } \hline
{\rm Fields} &  \left[ SU(4) \right] & SU(N_f)_L &SU(N_f)_R    \\ \hline \hline
p &\Yfund &{\overline{\Yfund}}&1   \\
\widetilde{p} & \overline{\Yfund}&1 & {\Yfund}  \\
{H} & 1&\Yfund &  \overline{\Yfund}    \\
 \hline \end{array} 
\]
\caption{Fermion and Higgs matter content and their transformations with respect to the Pati-Salam SU(4) gauge group.}
\label{PS}
\end{table}
 We have now $p^i_{\alpha,C}$  ( $\widetilde{p}^i_{\alpha,C}$) with $C=1,2,3$ representing the ordinary left-handed quarks (conjugated right spinors) while $C=4$ are the leptons, and therefor $C$ is the vector index of the Pati-Salam group $SU(4)$. The $B-L$ symmetry is automatically embedded as one of the generators of $SU(4)$  \cite{Mohapatra:1980qe}.  This description of the SM fields is, in practice, of book-keeping nature. To upgrade this model to a more realistic one Pati and Salam introduced several new  scalar degrees of freedom with the hope that one day there might be a more profound understanding of the origin behind the scalar sector.  Here, we will not duel with the specific details of the scalar potential and the pattern of chiral symmetry breaking. Therefore, we add only the minimum number of  matter fields allowing for such a possibility to manifest itself. We start by introducing the new complex scalars $\Phi_p$ ($\widetilde{\Phi}_{\widetilde{p}}$) transforming according to the fundamental (antifunamental) representation of the $ \left[ SU(4) \right] $ gauge group and fundamental (antifundamental) representation of the first (second) flavor group. 

At this point the spectrum of the theory is close to the {\it magnetic} gauge dual envisioned in \cite{Mojaza:2011rw}. The states to add are a  magnetic Weyl fermion $\lambda_m$ transforming according to the adjoint representation of $\left[ SU(4) \right]$  and a Weyl fermion $M$ transforming as the Higgs with respect to the non-abelian flavor group. Adding these states leads to the spectrum reported in Table~\ref{PSE}. 
\begin{table}[h]
\[ \begin{array}{|c| c | c c c  c| } \hline
{\rm Fields} &  \left[ SU(4) \right] & SU(N_f)_L &SU(N_f)_R & U(1)_p & U(1)_{AF} \\ \hline \hline
\lambda_m &{\rm Adj} & 1 &1 &~~0 &~~1\\
p &\Yfund &{\overline{\Yfund} }&1&~~\frac{2n_g - 4}{4}&-\frac{4}{2n_g} \\
\widetilde{p} & {\overline{\Yfund}}&1 &  {\Yfund}& -\frac{2n_g - 4}{4} &-\frac{4}{2n_g}\\
\Phi_p &\Yfund &{\overline{\Yfund} }&1&~~\frac{2n_g - 4}{4} &-\frac{2n_g -4}{2n_g}\\
\widetilde{\Phi}_{\widetilde{p}} & {\overline{\Yfund}}&1 & {\Yfund}& -\frac{2n_g - 4}{4}&-\frac{2n_g -4}{2n_g}  \\
{M} & 1&\Yfund &  \overline{\Yfund}& ~~0&-1+\frac{8}{2n_g}   \\
{H} & 1&\Yfund &  \overline{\Yfund}& ~~0 &\frac{8}{2n_g}  \\
 \hline \end{array} 
\]
\caption{The high-energy complete magnetic spectrum including the fields of the SM and their Pati-Salam extension.}
\label{PSE}
\end{table}
 We also make explicit the global symmetries of the new theory which are constituted by a new vector-like $U(1)_p$ and an axial one $U(1)_{AF}$ which is anomaly free. These global symmetries play a fundamental role via the 't Hooft anomaly conditions in order to identify the correct electric theory.  By determining the most general set of solutions to these conditions, together with requiring consistent flavor decoupling and involution at the level of the electric and magnetic gauge groups in  \cite{Mojaza:2011rw}  it was argued that the natural nonsupersymmetric electric dual theory is the one summarized in Table~\ref{dual}.
\begin{table}[t]
\[ \begin{array}{|c| c | c c c  c| } \hline
{\rm Fields} &  \left[ SU(2n_g - 4) \right] & SU(2n_g)_L &SU(2n_g)_R & U(1)_p & U(1)_{AF} \\ \hline \hline
\lambda &{\rm Adj} & 1 &1 &~~0 &~~1\\
P &\Yfund &{\Yfund }&1&~~1&-\frac{2n_g - 4}{2n_g} \\
\widetilde{P} & \overline{\Yfund}&1 &  \overline{\Yfund}& -1  &-\frac{2n_g - 4}{2n_g}\\
 \hline \end{array} 
\]
\caption{Electric dual of the magnetic Pati-Salam extension of the SM whose spectrum is summarized in 
}
\label{dual}
\end{table}
In \cite{Mojaza:2011rw} it was further shown that it is possible to construct all the singlet states of the magnetic theory as composites of the electric ones. The only state we need to add to the table of \cite{Mojaza:2011rw} is $H$ which corresponds naturally to the electric composite gauge singlet $P\lambda \lambda \widetilde{P}$. It is remarkable that the proposed electric dual theory does not contain any scalar degrees of freedom. 

{Since the magnetic extension of the SM mimics the Pati-Salam one I expect that part of the phenomenological analysis will resemble to the one already present in the literature (see for example \cite{Toorop:2010yh}) }. However, it is also clear that as we go higher in energy the magnetic description must deform itself into the electric one. In the same way the hadronic description dissolves itself into the quark and gluons one. This will require the discovery of copies of the SM fields at higher energies interacting via momentum dependent form factors.  

Following \cite{Mojaza:2011rw} the dual {\it electric} gauge group is therefore $SU(2n_g - 4) = SU(N_f  - 4)$. {}In order for the magnetic theory to be nonabelian we must have $2n_g - 4 \geq 2$ yielding the phenomenologically intriguing result that $n_g \geq 3$.  Of course, if $n_g=3$ the electric gauge group is $SU(2)$ and we expect an enhanced accidental global symmetry to occur i.e. $SU(2n_g)_L \times SU(2 n_g)_R \times U(1)_p  \subset SU(4 n_g)$, however if $n_g > 3 $ the electric theory has the same global symmetries of the magnetic one. Requiring the magnetic theory to remain asymptotically free we deduce the upper bound on the number of generations to be $6$ and therefore: 
\begin{equation}
3 \leq n_g \leq 6 \ .
\end{equation}
Interestingly duality, via  a Pati-Salam extension of the SM, hints to an elegant solution to the mystery of the phenomenological existence of, at least, three generations of quarks and leptons. 

Our construction predicts the existence of few more matter fields around the energy scale where the Pati-Salam extended color gauge group $SU(4)$ appears. We expect this scale to be above  the TeV scale. The reader will recognize that our magnetic spectrum resembles a supersymmetric one, however,  the magnetic theory is not supersymmetric since we do not invoke supersymmetric relations among its spectrum and couplings  \cite{Mojaza:2011rw}. 

The dual electric theory is expected to be strongly coupled at the energy scale where the magnetic one is weakly coupled explaining why the quarks, the leptons and the Higgs seem elementary.  The Pati-Salam like extension of the SM and our electric dual might not be unique and, in principle, an even more minimal magnetic extension of the SM with associated electric description could exist. A similar supersymmetric construction was attempted in \cite{Csaki:2011xn}.  

{ Some caveats  and further explanations are in order. Our results rely on the potential existence of gauge duals for nonsupersymmetric gauge theories featuring only fermionic matter whose dual involves both fermions and scalars. This type of duality has not yet been fully established for the theory presented here. There is, however, a known example in nature. This is ordinary QCD. The dual theory of QCD is the theory of hadrons which features composite fermions and scalars. The hadronic picture is weakly coupled at large N.  Therefore nonsupersymmetric gauge theories can have dual gauge theories in a fashion similar to QCD of supersymmetric examples. Arguably the first test for the possible existence of gauge duals is the demonstration that the dual passes the 't Hooft anomaly matching conditions. {}For the theory used here in \cite{Mojaza:2011rw} it has been exhibit the infinite set of solutions of the 't Hooft anomaly conditions for any number of flavors and colors.  We then required a number of universal (i.e. susy independent) constraints \cite{Mojaza:2011rw}.  

The introduced scalars in the magnetic theory are composite states of the electric degrees of freedom in the same way that Seiberg's postulated dual quarks are composite states of the electric quarks. A famous example is QCD itself where many spin zero states emerge and are  instrumental to explain the low energy physics  of QCD.   
 
\section{What will be discovered next?}
 
According to the minimal models of dynamical electroweak breaking we expect to see at the LHC the emergence of heavy technnirho and techniaxial states in the one to three TeV region with very small couplings to the SM fields \cite{Foadi:2007ue,Belyaev:2008yj,Andersen:2011yj}. These states will be discovered via Drell-Yan production and/or studied indirectly by investigating the composite Higgs production in association with a SM gauge boson \cite{Foadi:2007ue,Belyaev:2008yj,Andersen:2011yj}.  

If, on the other hand, composite dynamics emerges in the form of extreme compositeness we will observe a new zoo of composite particles some of which may carry colour. For this kind of extreme dynamics we do not yet have a firm prediction for the new scale of the theory although it is natural to expect it in the multi TeV range as well.

 \acknowledgements     
I thank Claudio Pica, Robert Shrock and Kimmo Tuominen for useful comments on the manuscript.  
CP$^3$-Origins is partially supported by the Danish National Research Foundation, grant number DNFR 90.

\end{document}